\documentclass[prl,twocolumn,showpacs,floatfix,amsmath,amsfonts]{revtex4}
\usepackage{amsmath}
\usepackage{epsfig}

\begin{document}

\title{Stabilized two-dimensional vector solitons}

\author{Gaspar D. \surname{Montesinos}}
\affiliation{Departamento de Matem\'aticas, Escuela T\'ecnica
Superior de Ingenieros Industriales, \\
Universidad de Castilla-La Mancha, 13071 Ciudad Real, Spain}

\author{V\'{\i}ctor M. \surname{P\'erez-Garc\'{\i}a}}
\affiliation{Departamento de Matem\'aticas, Escuela T\'ecnica
Superior de Ingenieros Industriales, \\
Universidad de Castilla-La Mancha, 13071 Ciudad Real, Spain}

\author{Humberto \surname{Michinel}}
\affiliation{\'Area de \'Optica, Facultade de Ciencias de Ourense,\\
Universidade de Vigo, As Lagoas s/n, Ourense, ES-32005 Spain.}

\date{\today}


\begin{abstract}
In this letter we introduce the concept of stabilized vector
solitons as nonlinear waves constructed by addition of mutually
incoherent Townes solitons that are stabilized under the effect of
a periodic modulation of the nonlinearity. We analyze the
stability of this new kind of structures and describe their
behavior and formation in Manakov-like interactions. Potential
applications of our results in Bose-Einstein condensation and
nonlinear optics are also discussed.
\end{abstract}

\pacs{42.65.Tg, 03.75. Fi, 05.45.Yv,}


\maketitle


Since the introduction of the concept of soliton as solitary water
waves with robust asymptotic behavior after mutual collisions,
many other physical systems have been found with similar dynamics,
always described by nonlinear wave equations \cite{Akh}. For
solitons of Nonlinear Schr\"odinger equations (NLSE), the main
interest in the early investigations was related with practical
applications in optical telecommunications, nowadays well
established \cite{KIV}. The recent interest on solitons in the
field of Bose-Einstein Condensation (BEC) in alkali gases with
negative scattering length \cite{bright1,bright2,bright3,bright4},
shows the timeliness of the topic and its central place in modern
physics.

Despite the success of the concept of soliton, these structures
arise mostly in 1+1-dimensional configurations. In the NLSE case
this is mainly due to the well known {\em collapse} property in
multi-dimensional scenarios \cite{Sulem}. In the Optical context,
{\em collapse} means that a laser beam with power higher than a
critique threshold, will be strongly self-focused to a singularity
when propagates in a Kerr-type nonlinear medium, whereas for lower
powers it will spread as it propagates. This behavior has also
been observed in experiments with matter waves
 \cite{collapse}.

Since collapse prevents the stability of multidimensional
``soliton bullets" in systems ruled by NLSE, a great effort has
been devoted to search for systems with stable solitary waves in
multidimensional configurations \cite{bullets}. A new way to
generate {\em stabilized} two-dimensional solitary waves has been
recently proposed for optical systems  \cite{Berge,IsaacLayered}.
The idea is to prevent collapse by using a spatial modulation of
the Kerr coefficient (the nonlinearity) of the optical material so
that the beam becomes collapsing and expanding in alternating
regions and is stabilized in average. The idea has been extended
to the field of matter waves in Refs. \cite{Ueda,pisaBoris}.
Finally, in Ref. \cite{Gaspar} some general results are provided.

In the present paper we will extend this analysis to the case of
mutually incoherent beams with unexpected and surprising results.
This is, to our knowledge, the first theoretical evidence of
two-dimensional stabilized vector solitons (SVS), a new kind of
nonlinear waves which can be constructed in two ways: by direct
combination of several Townes solitons or as a result of Manakov
interactions \cite{Manakov} between Townes solitons. In both
cases, the stabilization against collapse is obtained by the
effect of a peridodic modulation of the nonlinearity.


{\em The model.-} Let us consider a $n$-component system modelled
by equations of the type
\begin{equation}
\label{Manakoveqs}
i \frac{\partial u_j}{\partial t}  =
- {1 \over 2} \Delta u_j + g(t) \left(\sum_{k=1}^n a_{jk} |u_k|^2\right)u_j,
 \end{equation}
where $j = 1,\ldots, n$, $u_j: \mathbb{R}^+\times
\mathbb{R}^2\rightarrow \mathbb{C}, \Delta =
\partial^2/\partial x^2 + \partial^2/\partial y^2, a_{jk} \in
\mathbb{R}$ are the nonlinear coupling coefficients and $g(t)$ is
a periodic function accounting for the modulation of the
nonlinearity.

Eqs. (\ref{Manakoveqs}) are the natural extension of the Manakov
system \cite{Manakov} to two transverse dimensions and an
arbitrary number of components. In Optics, for spatial solitons
$t$ is the propagation coordinate and $u_j$ are $n$ mutually
incoherent beams. One-dimensional Manakov-type models have been
extensively studied in nonlinear optics, mainly due to the
potential applications of Manakov solitons in the design of
all-optical computing devices \cite{todos}. In BEC these equations
(with an additional trapping term) describe the dynamics of
multicomponent condensates, $u_{j}$ being the wavefunctions for
each of the atomic species involved \cite{PRLdual,Nature}.

In the scalar case ($n=1$), it is well known that, if $g$ is
constant, there is a stationary radially symmetric solution of Eq.
\eqref{Manakoveqs} (the so-called Townes soliton): $u({\mathbf r},
t) = \Phi(r)e^{i\lambda t}$. This solution is {\em unstable}
meaning that the value of the norm $N^{1/2} = (\int
|\Phi|^2)^{1/2}$ is critical in the sense that any generic slight
perturbation of the initial condition will yield to collapse or
spreading of the distribution. We must also notice that, due to
the scaling invariance of the cubic NLS, a family of Townes
solitons can be generated  by making $\Phi_\lambda(r,t)= \lambda
\Phi(\lambda r, \lambda^2t)$.

It has been shown in \cite{Gaspar} that an adequate modulation of
the nonlinearity, will asymptotically stabilize a Townes soliton
yielding to a rapidly oscillating {\em stabilized} Townes soliton
(STS), which we refer hereatfer as $\Phi_S$. In this paper, we
take  $g(t) = g_0 + g_1 \cos \Omega t$ but we expect that most of
our results with similar periodic functions will be qualitatively
the same \cite{Gaspar}.

{\em Stabilized Vector Solitons (SVS).-} For a given set of
parameters $a_{jk}$ it is possible to use stabilized Townes
solitons to build explicit solutions of Eqs. \eqref{Manakoveqs}.
These solutions are constructed by taking
$u_j=\Phi_{S_j}\equiv\alpha_j \Phi_S, j=1,\ldots, n$ for any set
of coefficients $\alpha_j$ satisfying
\begin{equation}
\label{coefs} a_{j1} \alpha_1^2 + ... + a_{jn}\alpha_n^2 = 1,
j=1,\ldots, n.
\end{equation}


\begin{figure}
\epsfig{file=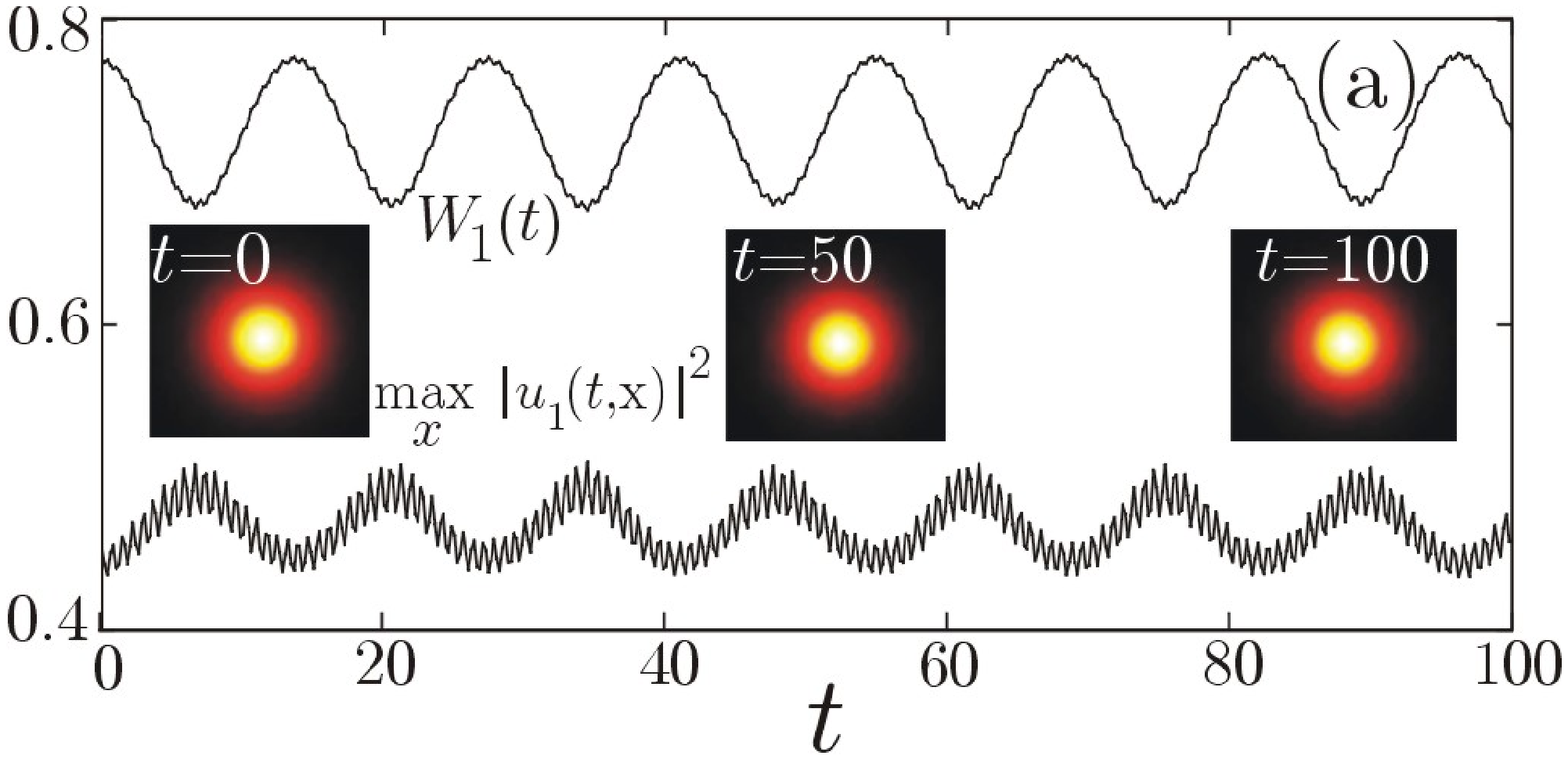,width=\columnwidth}
\epsfig{file=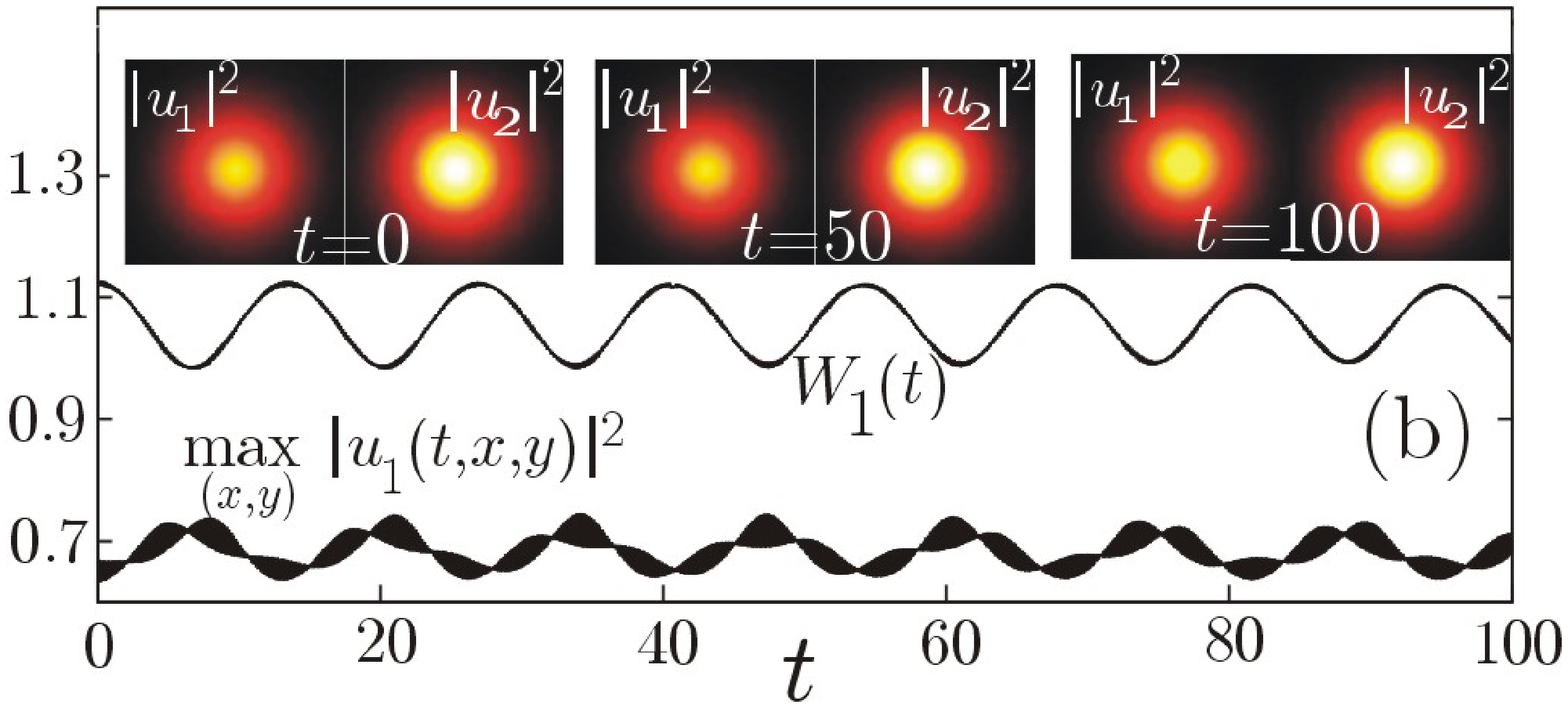,width=\columnwidth} \caption{[Color
online] (a) Stabilized vector solitons for $\alpha_1 = \alpha_2 =
1/\sqrt{2}$ (the evolution for $u_2$, not shown here, is very
similar). Shown are the time evolution of the width $W_1 =
(\int_{\mathbb{R}^2} (x^2+y^2) |u_1|^2)^{1/2}$ and amplitude
$\max_{x\in \mathbb{R}^2} |u_1(x,t)|$. The insets show
 pseudocolor plots of $|u_1(t,x,y)|^2$ for different times. (b)
Same as (a) but for $\alpha_1 = 1/\sqrt{3}, \alpha_2 =
\sqrt{2/3}$. For both cases $g(t) = -2\pi+8\pi\cos(40 t)$.
\label{prima}}
\end{figure}

It is not obvious that these {\em new} solutions will be
stabilized by a periodic modulation $g(t)$ of the nonlinearity.
Writing $u_j = \Phi_{S_j} + \delta_j$ the equations for $\delta_j$
contain cross-modulation terms which could lead to growth of these
small perturbations. To test that the stabilization is possible in
a wide range of configurations, we have considered several
important examples. First, we have studied the most relevant case
$n=2$ and integrated numerically Eqs. \eqref{Manakoveqs} with
different initial data of the form $u_j = \alpha_j \Phi_S$
satisfying (\ref{coefs}) and found that these new vector solitons
remain stabilized as shown in Fig. \ref{prima}. From now on we
will name these structures as Stabilized Vector Solitons (SVS). We
will see below how they emerge in collisions of mutually
incoherent STSs, which correspond to the so-called Manakov
interactions \cite{Manakov}.

We have studied other situations such as a symmetric superposition
of four STS ($n=4$) with $\alpha_j = 1/\sqrt{4}$ and found similar
results. Thus, these structures exist in a wide range of
parameters and configurations.


{\em Manakov interactions of STS}.- Depending on the mutual
velocity of the two interacting STS, we have divided the regime of
collisions in two different ones: fast and slow collisions. As we
will see below, one of the main results of our work is the
possibility of obtaining SVS after {\em slow} collisions of STS.

\begin{figure}
\epsfig{file=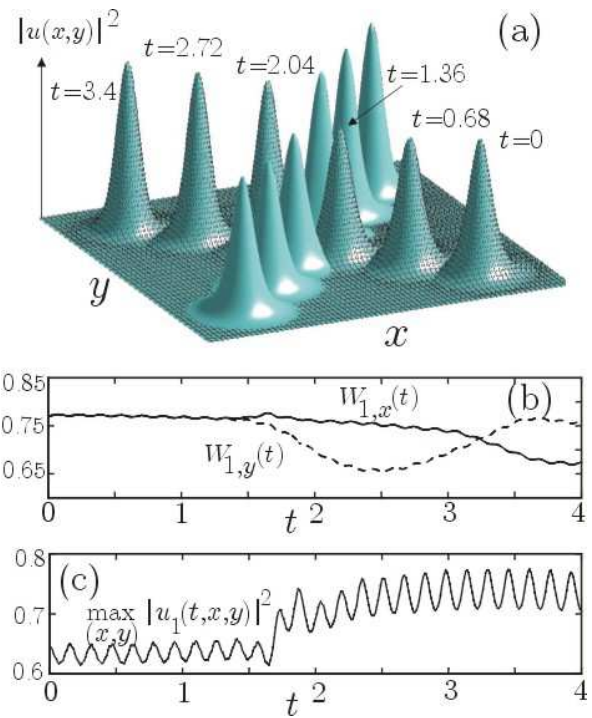,width=0.9\columnwidth} \caption{[Color
online] Fast collisions of stabilized Townes solitons. Initial
data are $u_1(0,\mathbf{r}) =
e^{i\mathbf{v}_1\mathbf{r}}\Phi(\left|\mathbf{r}+\mathbf{r}_1\right|)$,
$u_2(0,\mathbf{r}) =
e^{i\mathbf{v}_2\mathbf{r}}\Phi(\left|\mathbf{r}+\mathbf{r}_2\right|)$
with $\mathbf{v}_1 = (5/\sqrt{2}, 5/\sqrt{2})$, $\mathbf{v}_2 =
(-5/\sqrt{2}, 5/\sqrt{2})$ and $\mathbf{r}_1 = (-6,-6)$,
$\mathbf{r}_2 = (6,-6)$. (a) Surface plots of $|u_1|^2$ and
$|u_2|^2$ for different times. (b) Evolution of the widths
$W_{1,x}(t) = (\int \left(x-<x>\right)^2 |u_1(x,y,t)|^2)^{1/2}$,
$W_{1,y}(t) = (\int \left(y-<y>\right)^2 |u_1(x,y,t)|^2)^{1/2}$.
(c) Evolution of the maximum amplitude $\max_{(x,y) \in
\mathbb{R}^2} \left|u_1(x,y,t)\right|$. \label{segunda}}
\end{figure}

First we have studied collisions of ``fast" STSs after which the
solitons emerge with only moderate modifications of their
amplitude and width as is shown in Fig. \ref{segunda}. It can be
seen [Fig. \ref{segunda}(b)] that during the collision, the
soliton becomes spatially asymmetric. An internal asymmetric
breathing mode of small amplitude is excited which decays at
longer times (not shown in the figure) to the ``normal" symmetric
breathing mode shown by STSs.


These behaviors can be accounted for by a finite-dimensional
reduction  of Eqs. \eqref{Manakoveqs} by means of the
time-dependent variational approach. Notice that Eqs.
\eqref{Manakoveqs} can be obtained from the Lagrangian density
\begin{multline}
\mathcal{L} = \frac{i}{2} \left(u_1 \frac{\partial u_1^*}{\partial
t} + u_2 \frac{\partial u_2^*}{\partial t} + \text{h.c.} \right) +
\frac{1}{2}|\nabla u_1|^2 +
\frac{1}{2}|\nabla u_2|^2  \nonumber \\
+  \frac{g(t)}{2} \left(a_{11}|u_1|^4 + 2a_{12}|u_1|^2|u_2|^2+
a_{22}|u_2|^4\right).
\end{multline}
We choose a simple ansatz accounting for head-on symmetric
collisions of equal stabilized solitons moving with opposite
speeds and centered on $(-\ell,0)$ and $(\ell,0)$
\begin{subequations}\label{ansatz}
\begin{eqnarray} u_1 & = & A
e^{-(x-\ell)^2/2\omega_x^2-y^2/2\omega_y^2+i\beta_xx^2+i\beta_yy^2}e^{ivx},
\\
u_2 & = & A
e^{-(x+\ell)^2/2\omega_x^2-y^2/2\omega_y^2+i\beta_xx^2+i\beta_yy^2}e^{-ivx}.
\end{eqnarray}
\end{subequations}
Although gaussians do not have the right asymptotic decay as STSs,
our choice simplifies the calculations and is enough for our
present objectives. The standard variational method \cite{Boris}
leads to the equations (for $a_{jk}=1$)
\begin{subequations}
\label{variacional}
\begin{eqnarray}
\ddot{\ell} & = & \ell\frac{N g(t)}{\pi w_x^3 w_y}
e^{-2\ell^2/w_x^2}, \\ \label{wa} \ddot{w}_x & = &
\frac{1}{\omega_x^3}+\frac{Ng(t)}{2\pi\omega_x^2 \omega_y}\left [
1+e^{-2\ell^2/w_x^2}\left (
1-\frac{4\ell^2}{\omega_x^2}\right)\right],\\ \label{wb}
\ddot{w}_y & = & \frac{1}{w_y^3} + \frac{Ng(t)}{2\pi w_x
w_y^2}\left(1+e^{-2\ell^2/w_x^2}\right),
\end{eqnarray}
\end{subequations}
together with the complementary relations $\beta_j =
\dot{w}_j/2\omega_j$, $(j=x,y)$, $v = \dot{\ell} - 2\ell \beta_x$,
and the conservation law $N(t) = \pi |A|^2 \omega_x \omega_y = \pi
|A(0)|^2 \omega_x(0) \omega_y(0)$. The different terms in Eqs.
\eqref{variacional} account for the phenomenology shown in Fig.
\ref{segunda} and other ``fast collisions" studied. For example,
they contain the asymmetric interaction (notice the differences
between Eqs. \eqref{wa} and \eqref{wb}) due to the fact that both
solitons approach along the $x$ axis and thus become more
elongated along that direction as seen in Fig. \ref{segunda}(b).
We have numerically integrated Eqs. \eqref{variacional} for fast
collisions taking as initial data stabilized gaussian functions
\cite{IsaacLayered,Gaspar} and have found results similar to those
shown in Fig. \ref{segunda}.


\begin{figure}
\epsfig{file=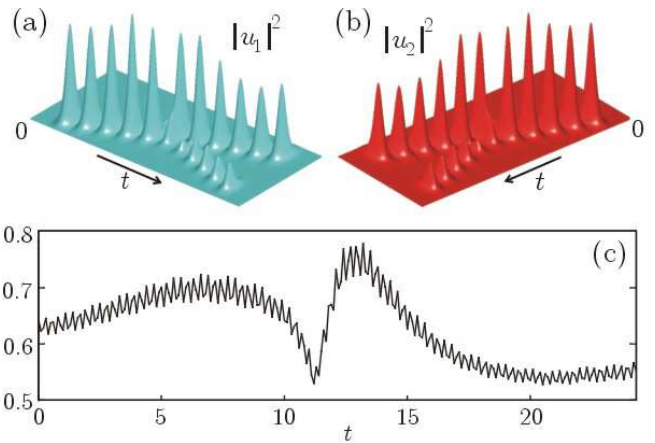,width=\columnwidth} \caption{[Color
online] Head-on collisions of STSs for initial data $v_1 =-v_2 =
0.3.$ (a) Surface plots of $\left|u_1\right|^2$, (b) surface plots
of $\left|u_2\right|^2$. (c) Evolution of the maximum amplitude
$\max_{(x,y) \in \mathbb{R}^2} \left|u_1(x,y,t)\right|$
\label{tertia}}
\end{figure}

\begin{figure}
\epsfig{file=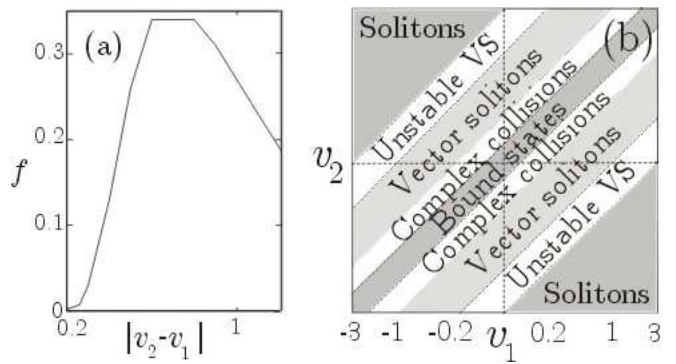,width=\columnwidth}
\caption{Asymptotic behavior after head-on collisions of STSs of
the form $u_1 = \Phi(|\mathbf{r}-\mathbf{\ell}|)e^{iv_1x}$, $u_2 =
\Phi(|\mathbf{r}+\mathbf{\ell}|)e^{iv_2x}$ and large enough $\ell$
$(\sim 4)$. (a) Regimes of behavior as functions of $v_1, v_2$.
(b) Quotient ($f$) of the squared amplitudes of the small and
large peaks which are generated after the collision when a VS is
formed [see Fig. \ref{tertia}] for the regime of speeds in the
range $0.2 \leq |v_2-v_1| \leq 1.2$. \label{plano}}
\end{figure}

The regime of slow collisions is in the range $|v_2-v_1| \sim 3$.
In this case the collisions of STSs lead to the formation of two
vector solitons as shown in Fig. \ref{tertia}(a,b). It is
remarkable and one of the main results of the paper that the
collision mechanism allows the complex coherent rearrangement
necessary for the formation of the vector solitons. The fraction
of ``mass" interchanged by the incoming solitons is a function of
the only relevant parameter for direct collisions $|v_2-v_1|$ (due
to the Galilean invariance) as shown in Fig. \ref{plano}(a). In
the range $0.2<|v_2-v_1|<3$ we observe formation of two vector
solitons which seem to be either unstable or performing high
amplitude oscillations for higher speeds and stable in the lower
range of speeds (approximately $0.2<|v_2-v_1|<1.2$). If the speed
is decreased further we observe two outgoing vector solitons with
complex transient dynamics and nontrivial dependence of the
fraction transferred as a function of $|v_2-v_1|$.


\begin{figure}
\epsfig{file=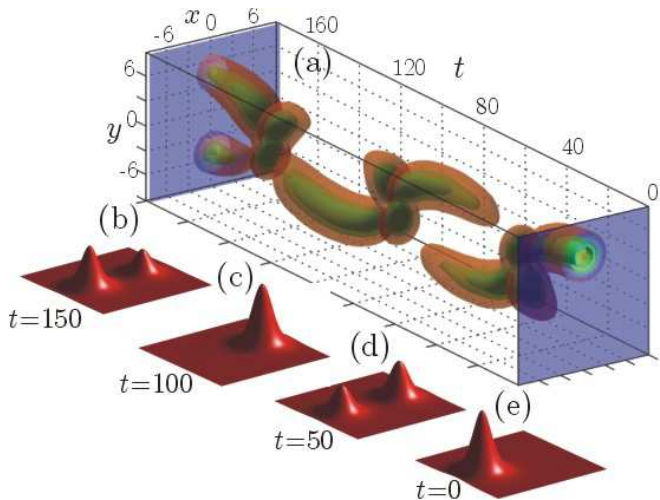,width=\columnwidth} \caption{[Color
online] Oscillations after a collision of STSs with $v_1 = v_2 =
0$. (a) Isosurface plot of $|u_1(t,x,y)|^2$ (shownare isosurfaces
corresponding to 0.05, 0.15 and 0.25), (b,c,d,e) Surface plots of
$|u_1(t,x,y)|^2$ for (b) $t=0$, (c) $t = 50$, (d) $t = 100$, (e)
$t=150$. The corresponding evolution for $u_2$ is symmetric with
respect to the $y$ axis. \label{cuarta}}
\end{figure}

Finally, if the initial speed of the colliding solitons is very
small or zero, we have observed a quasi-bound state of two SVSs
which shows several recurrent collisions as shown in Fig.
\ref{cuarta}. From our simulations we cannot conclude if this is a
true bound state or it finally decays to vector solitons. In Fig.
\ref{plano}(b) we summarize the results of our numerical
exploration of STS collisions.

We want to point out that Eqs. \eqref{variacional} provide a
reasonably good description of the phenomena described here as far
as the ansatz given by Eqs. (\ref{ansatz}) can describe these
complex dynamical behaviors. An example: for very low speeds the
variational equations predict the formation of an oscillating
bound state of two STSs. Although this is not the real behavior (a
bound state of two SVSs is formed) we get bound states.

The formation of vector solitons from stabilized scalar solitons
is a nontrivial phenomenon since there is a delicate balance of
both components which must be satisfied in order to avoid
destabilization either to collapse or expansion of these
structures. It is curious that the system is able to interchange
just the right amount of energy to keep both solitons bounded. In
fact, the collision mechanism described here can be seen as a way
to generate appropriate stabilized vector solitons up from STSs
which could be otherwise difficult to obtain. This is another
proof of the structural stability of these new physical objects.

It is also remarkable that no collapse phenomena is observed in
our simulations, rather instead most of the collisions observed
lead to remarkably robust scalar or vector solitons. This is very
different from what happens in coherent collisions of STSs which
lead to collapse.


\emph{Conclusions.-} In this paper we have described a new type of
vector solitons, the stabilized vector solitons (SVS). We have
studied their stability and shown how they arise in Manakov
collisions of STSs. Other phenomena seen in collisions of STSs
have been described and analyzed such as ``quasi"-bound state
formation and excitation of asymmetric oscillation modes.

We must stress that the limit $n\rightarrow\infty$ of our model
could be used to study nonlinear propagation of totally incoherent
light. In BEC a detailed investigation of the above system could
have implications in the study of decoherence effects in BECs.


G. D. M. and V. M. P-G. are partially supported by Ministerio de
Ciencia y Tecnolog\'{\i}a (MCyT) under grants BFM2000-0521 and
BFM2003-02832 and Consejer\'{\i}a de Ciencia y Tecnolog\'{\i}a de
la Junta de Comunidades de Castilla-La Mancha under grant
PAC-02-002. G. D. M. acknowledges support from grant AP2001-0535
from MECD. H. M. is partially supported by MCyT under grant
TIC-2000-1105-C03-01.


\end{document}